\documentclass[twocolumn,showpacs,preprintnumbers,amsmath,amssymb]{revtex4}
\usepackage{graphicx}
\usepackage{dcolumn}
\usepackage{bm}

\usepackage{epsfig}
\newcommand{\beq}{\begin{equation}}
\newcommand{\eeq}{\vspace{0cm} \end{equation}}
\newcommand{\beqq}{\setlength\arraycolsep{2pt}\begin{eqnarray}}
\newcommand{\eeqq}{\vspace{0cm} \end{eqnarray}}

\begin{document}

\title{CDM Accelerating Cosmology as an Alternative to $\Lambda$CDM model}

\author{J. A. S. Lima$^{1,2}$}\email{limajas@astro.iag.usp.br}

\author{J. F. Jesus$^{1}$}\email{jfernando@astro.iag.usp.br}

\author{F. A. Oliveira$^{1}$}\email{foliveira@astro.iag.usp.br}

\affiliation{$^{1}$Departamento de Astronomia, Universidade de S\~ao Paulo\\
Rua do Mat\~ao, 1226, 05508-900, S\~ao Paulo, SP, Brazil}

\affiliation{$^{2}$Center for Cosmology and Astro-Particle Physics, The Ohio State University, \\  
191 West Woodruff Avenue, Columbus, OH 43210, USA}

\begin{abstract}\begin{center}{\bf Abstract}\end{center}
A new accelerating cosmology  driven only by baryons plus cold dark matter (CDM) is proposed in the framework of general relativity. In this model the present accelerating stage of the Universe is powered by the negative pressure describing the gravitationally-induced particle production of cold dark matter particles.  This kind of scenario has only one free parameter and the differential equation governing the evolution of the scale factor is exactly the same of the $\Lambda$CDM model. For a  spatially flat Universe, as predicted by inflation ($\Omega_{dm}+\Omega_{baryon}=1$), it is found that the effectively observed  matter density parameter is  $\Omega_{meff} = 1- \alpha$,   where $\alpha$ is the constant parameter specifying the CDM particle creation rate. The supernovae test based on the Union data (2008) requires $\alpha\sim 0.71$ so that $\Omega_{meff} \sim 0.29$  as independently derived  from weak gravitational lensing, the large scale structure and other complementary  observations. 

\end{abstract}


\maketitle

\section{Introduction}

It is well known that observations  from Supernovae Type  Ia (SNeIa) provide strong evidence for
an expanding accelerating Universe \cite{Riess07,Union08}. In relativistic cosmology, such a phenomenon is usually explained by the existence of a new dark component (in addition to cold dark matter), an exotic fluid endowed with negative pressure \cite{review}.  

Many candidates for dark energy have been proposed in the literature \cite{decaying,XM,SF,CGas}, among them: (i) A cosmological constant ($\Lambda$), (ii) a decaying vacuum energy density or $\Lambda(t)$-term,  (iii) a relic scalar field slowly rolling down its potential, (iv) the
``X-matter'',  an extra component characterized by equation
of state $p_x=\omega \rho_x$, where $\omega$ may be constant or a redshift dependent function, (iv) a Chaplygin-type
gas whose equation of state is $p=-A/\rho^{\gamma}$, where 
A and $\gamma$ are positive parameters.  All these models explain the accelerating stage, and, as such, the space parameter of the basic observational quantities is rather degenerate. Nowadays, the most economical explanation is provided by the flat $\Lambda$CDM model which has only dynamic free parameter, namely, the vacuum energy density. It seems to be consistent with all the available observations provided that the vacuum energy density is fine tuned to fit the data ($\Omega_{\Lambda} \sim 0.7$).  However, even considering that the addition of these fields explain the late time accelerating stage  and other complementary observations \cite{CMB,Clusters}, the need of (yet to be observed) dark energy  component with  unusual properties is certainly a severe hindrance.

In general relativistic cosmology, the presence of a negative pressure is the key ingredient to 
accelerate the expansion. In particular, this means  that cosmological models dominated by pressureless fluid like a CDM component 
expands in a decelerating way. However, as first discussed by Prigogine and coworkers \cite{Prigogine} and somewhat 
clarified by Calv\~{a}o and collaborators \cite{LCW}  through a manifestly covariant formulation, the  matter creation process at the expense of the gravitational field is also macroscopically described by a negative pressure. Later on, it was 
also demonstrated that the matter creation is an irreversible process 
completely  different from the bulk viscosity description \cite{LG92} 
originally proposed by Zeldovich \cite{Zeld70} 
to avoid the singularity, as well as to describe phenomenologically the 
emergence  of particles in the begin of the Universe evolution early (see also \cite{SLC02} 
for a more complete discussion comparing particle creation and bulk viscosity).

Microscopically, the  gravitationally induced particle creation mechanism has also been discussed by many authors \cite{Parker,BirrellD}. 
A non-stationary gravitational background influences quantum fields in such a way that the frequency 
becomes time-dependent. In the case of a flat Friedmann-Robertson-Walker (FRW) spacetime described in conformal time coordinates, the key result is that the scalar
field obeys the same equation of motion as a massive scalar field in Minkowski spacetime,
except that the effective mass becomes time dependent (the dispersion relation in the FRW
metric involves the scale factor and its second derivative). When the field is quantized, this
leads to particle creation, with the energy for newly created particles being supplied by the
classical, time-varying gravitational background. 

In this context, we are proposing here a new flat cosmological scenario where the cosmic acceleration is 
powered uniquely by the  creation of cold dark matter particles.  It will be assumed that the CDM particles are described by a real scalar field  so that only particle creation takes place because in this case it is its own antiparticle. The model can be seen as a workable alternative to the
cosmic concordance cosmology because it has only one free parameter and the equation of motion is exactly the same of the $\Lambda$CDM model.  As we shall see, in the case of a  spatially flat Universe ($\Omega_{dm}+\Omega_{bar}=1$), the effectively observed  matter density parameter is  $\Omega_{eff} = 1- \alpha$,   where $\alpha$ is the constant parameter defining the creation rate. The supernovae test requires the central value $\alpha\sim 0.71$ so that $\Omega_{eff} \sim 0.29$  in accordance with the large scale structure and other complementary  observations.  

\section{Cosmic Dynamics with creation of CDM Particles}

For the sake of generality, let us  start with the homogeneous and isotropic FRW line element
\begin{equation}
\label{line_elem}
  ds^2 = dt^2 - R^{2}(t) \left(\frac{dr^2}{1-k r^2} + r^2 d\theta^2+
      r^2{\rm sin}^{2}\theta d \phi^2\right),
\end{equation}
where $R$ is the scale factor and $k= 0, \pm 1$ is the curvature 
parameter. Throughout we use units such that $c=1$.

In that background, the nontrivial cosmological equations for the mixture of radiation, baryons and cold dark matter (with 
creation of dark matter particles), and the energy conservation laws for each component takes the following form \cite{ZP2,LG92,LGA96,LSS08}

\begin{equation}
\label{fried}
    8\pi G (\rho_{rad} + \rho_{bar}  + \rho_{dm}) = 3 \frac{\dot{R}^2}{R^2} + 3 \frac{k}{R^2},
\end{equation}

\begin{equation}
\label{frw_p}
   8\pi G (p_{rad} + p_{c}) =  -2 \frac{\ddot{R}}{R} - \frac{\dot{R}^2}{R^2} -
	\frac{k}{R^2},
\end{equation}
\begin{equation}\label{energy}
      \frac{\dot{\rho}_{rad}}{{\rho}_{rad}} + 4 \frac{\dot{R}}{R} = 0, \,\,\,\,\,\, \, \,\,\,\,\, \frac{\dot{\rho}_{bar}}{{\rho}_{bar}} + 3 \frac{\dot{R}}{R} = 0,
\end{equation}
and 
\begin{equation}
\label{ConsDM}
\frac{\dot{\rho}_{dm}}{{\rho}_{dm}} + 3 \frac{\dot{R}}{R} = \Gamma.
\end{equation}
In the above expressions,  an overdot means time derivative and   $\rho_{rad}$, $\rho_{bar}$ and $\rho_{dm}$, are the  radiation, baryonic and dark matter energy densities, whereas $p_{rad}$ and $p_{c}$,  denote the radiation and creation pressure, respectively. The quantity $\Gamma$ with dimension of  $(time)^{-1}$ is the creation rate of the cold dark matter. As should be expected, the creation pressure  is defined in terms of the creation rate and other physical  quantities. In the case of adiabatic creation of dark matter, it is given by \cite{Prigogine,LCW,LG92,ZP2,LGA96,SLC02,LSS08}
\begin{equation}
\label{CP}
    p_{c} = -\frac{\rho_{dm} \Gamma}{3H},
\end{equation}
where $H = {\dot {R}}/R$ is the Hubble parameter. 

The above expressions show how the matter creation rate,  $\Gamma$, modifies the 
evolution of the scale factor and the density of the cold dark matter as compared to the case with no creation. 
By taking  $\Gamma=0$ the above set reduces to the  differential equations governing the evolution of  radiation plus a pressureless  fluid mixture (baryons + CDM), as given by the FRW type cosmologies. 
 
\section {Creation Cold Dark Matter  (CCDM) Cosmology }

Let us now propose a new class of models  defined by the choice for the particle creation rate, 
$\Gamma$.  In principle, the most natural choice would be a particle creation rate which favors 
no epoch in the evolution of the Universe, that is,  $\Gamma \propto H$, where $H$ 
is the Hubble parameter.  

Recently, we have investigated  CCDM scenarios with $\Gamma=3\beta H$,  
where $\beta$ is a time-dependent  dimensionless parameter \cite{LSS08,SSL09}. 
In the first paper \cite{LSS08}, we have demonstrated that CCDM models solve the age problem and are generically capable of accounting for 
the SNIa observations. In the subsequent paper \cite{SSL09}, we have included baryons  
and tested the evolution of such models at high redshift 
using the constraint on $z_{eq}$, the redshift of the epoch of matter 
- radiation equality, provided by the WMAP constraints on the  early 
Integrated Sachs-Wolfe effect (ISW).  Such a comparison revealed 
a tension between the high redshift CMB constraint on $z_{eq}$ and 
that which follows from the low redshift SNIa data, thereby challenging the 
viability of that class of models. A minor caveat is related to the relative mathematical difficulty faced when the baryon component was introduced in the  CCDM cosmology proposed  by Lima, Silva and Santos \cite{LSS08}. Actually, in the most interesting scenario discussed by Steigman et al. \cite{SSL09}, the comparison with the observations became possible only after to expand the Hubble parameter function at low and high redshifts. 

On the other hand, the most essential difficulty of such CCDM models comes from the fact that all of them are flat ($\Omega_{dm} + \Omega_{bar}= 1$), but it is not clear how they can account for the clusters data which are consistently pointing to $\Omega_{dm}+ \Omega_{bar} \sim 0.3$ from a large set of observations \cite{Clusters}.  In particular, this means that the following question is challenging these CCDM scenarios. How the matter creation rate affects the present amount of matter  so that the effectively measured matter density parameter is close to that one obtained from the available observations? 

In what follows, we show that all these shortcomings can be solved at once through a reasonable choice of $\Gamma$. First we observe that the acceleration is a low redshift phenomenon, that is, it must be suppressed during the radiation phase. This may be obtained by considering that the  $\beta(t)$ function  is inversely proportional to the CDM  energy density itself. Since such a quantity is dimensionless it may depend on some ratio involving the dark matter energy density. To be more specific, let us  consider the following creation rate:
\begin{equation}
\label{Gamma} 
\Gamma=3\alpha \left(\frac{\rho_{co}}{\rho_{dm}}\right)H,
\end{equation}
where $\alpha$ is a constant parameter, ${\rho_{co}}$ is the present day value of the critical density, and the factor 3 has been maintained for mathematical convenience. 

Now, by inserting the above expression in the energy conservation for dark matter as given by ($\ref{ConsDM}$) one obtains
\begin{equation}
 \dot{\rho}_{dm} + 3H\rho_{dm}=\Gamma \rho_{dm}\equiv 3\alpha {\rho_{co}}H,
\end{equation}
which can be readily integrated to give  a solution for $\rho_{dm}$
\begin{equation}
\label{rhodm}
\rho_{dm} = (\rho_{dmo} - \alpha\rho_{co})\left({\frac{R_0}{R}}\right)^{3} + \alpha\rho_{co} 
\end{equation}
or, in terms of the redshift, $1+z=R_0/R$,

\begin{equation}
\rho_{dm} = (\rho_{dmo} - \alpha\rho_{co})(1 + z)^{3} + \alpha\rho_{co}. 
\end{equation}
Since the  solution of the energy conservation laws for radiation and baryons are the usual ones, namely, $\rho_{rad}=\rho_{rad0}(1+z)^4$, $\rho_b=\rho_{bar0}(1+z)^3$, by inserting these expressions into Friedmann equation (\ref{fried}), we arrive to (from now on we neglect the radiation fluid since in the present paper we test only the matter stage)
\begin{equation}
\label{Hz}
 \left(\frac{H}{H_0}\right)^2=(\Omega_{m}-\alpha)(1+z)^3+\alpha+(1-\Omega_{m})(1+z)^2,
\end{equation}
where we have defined 
\begin{equation}
 \Omega_m \equiv \Omega_{dm} + \Omega_{bar}, 
\end{equation}
and used the normalization condition to fix $\Omega_k=1-\Omega_{m}$. The similarity of the above expression with the one of the $\Lambda$CDM model is astonishing because it has been obtained by considering just one dark component. Actually, the Hubble parameter for a $\Lambda$CDM model reads:
\begin{equation}
\left(\frac {H_{\Lambda CDM}}{H_0}\right)^2 = \Omega_m(1+z)^3 + \Omega_{\Lambda} + (1 -\Omega_m -\Omega_{\Lambda})(1+z)^2
\end{equation}
One may see that the models have the same Hubble parameter $H(z)$, with $\alpha$ playing the dynamical role of $\Omega_{\Lambda}$ and $\Omega_m$ now being replaced by $\Omega_m-\alpha$. Such a map becomes even more clear by defining an `effective' matter density parameter, as $\Omega_{meff}\equiv\Omega_m-\alpha$ and inserting the result into the expression (\ref{Hz}) for the CCDM model.

This intriguing equivalence can also be seen even more directly through the evolution equation of the scale factor function.  As one may check, by inserting the expression of the creation pressure $p_c$ in the second Friedman equation we obtain:

\begin{equation}
\label{evolR}
2R{\ddot R}+ {\dot{R}}^2 + k - 3\alpha{H_0}^{2} R^{2}  = 0,  
\end{equation} 
which should be compared to: 
\begin{equation}
\label{evolRLCDM}
2R{\ddot R}+ {\dot{R}}^2 + k - {\Lambda}R^{2}  = 0,  
\end{equation}
provided by the $\Lambda$CDM model. The above equations imply that the models will have  the same dynamic behavior when we identify the creation parameter by the expression $\alpha = {\Lambda}/3{H_0}^{2} \equiv \Omega_{\Lambda}$, which is exactly the same result derived earlier with basis on the Hubble parameter, $H(z)$.

On the other hand,  although considering that $CCDM$ and $\Lambda$CDM models have the same dynamic history, such cosmologies are based on different starting hypothesis, and, as such, they  can be differentiated by the present observations. Mathematically, this happens  because of the special role played by the $\alpha$ parameter in the basic equations of the CCDM model.  In particular, the positiviness of the dark matter density (and Hubble parameter) at high redshifts implies that the creation parameter must satisfy the  constraint, $\alpha \leq \Omega_m$, a condition  absent in the $\Lambda$CDM case (see Eqs. \ref{rhodm} and \ref{Hz}). Further, the redshift dependence of the contribution involving the $\alpha$ parameter, namely, $\alpha(1 - (1+z)^{3})$ must modify slightly the predictions involving the evolution of small perturbations and the structure formation problem.
 
\begin{figure*}
\centerline{\epsfig{file=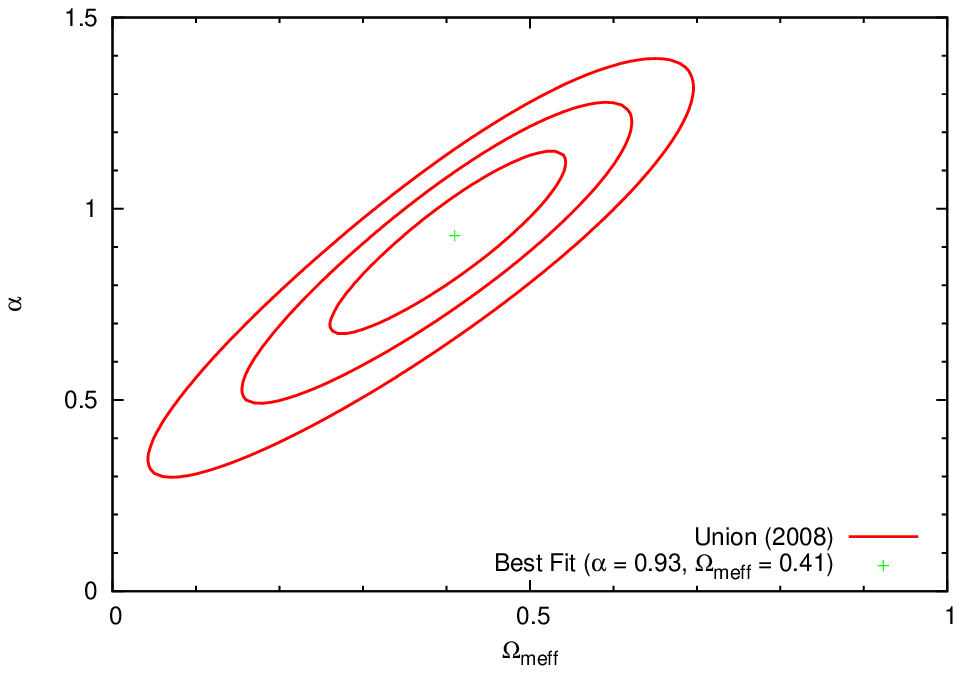,width=80mm, height=58mm}
\epsfig{figure=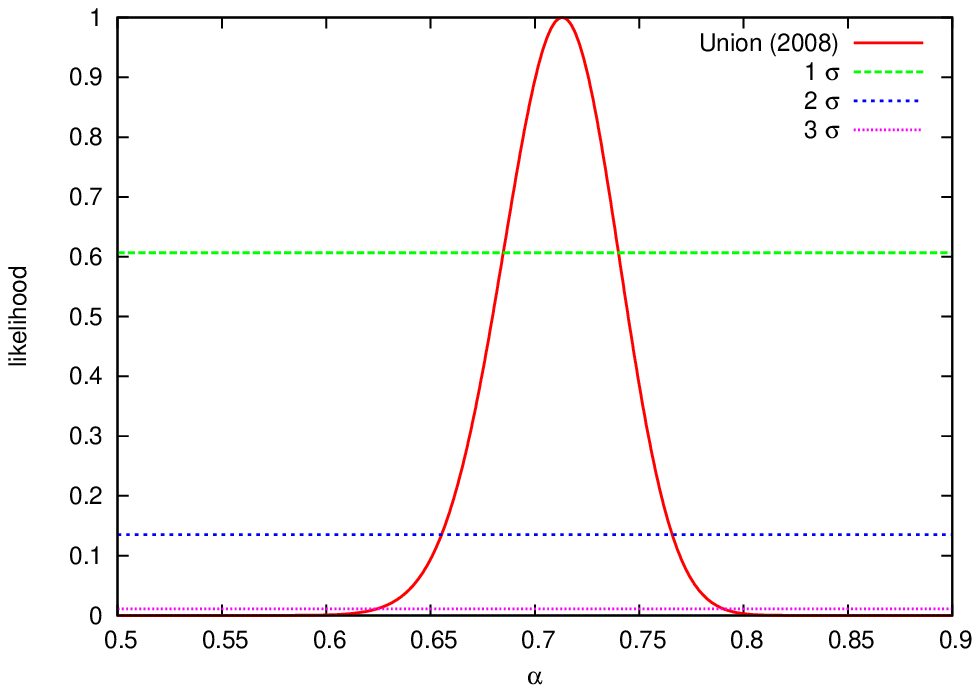, width=80mm, height=58mm}
\hskip
0.1in} \caption{{\bf {a)}} The $\alpha$ - $\Omega_{meff}$ plane predicted by the CCDM model based on 307 Supernovae data (Union) from Kowalski et al. (2008), showing contours of 68.3\%, 95.4\% and 99.7\% c.l. for two free parameters. Like in $\Lambda$CDM model these data also favors a closed Universe with creation of CDM particles ($\Omega_m = 1.34^{+0.54}_{-0.68} (2\sigma)$). {\bf{b)}} The likelihood for $\alpha$ in the flat CCDM scenario.} \label{fig1}
\end{figure*}

If spatial flatness is to be imposed, as predicted by inflation and suggested from CMB data, we would have $\Omega_m=1$, thus Eq. (\ref{Hz}) reduces to:
\begin{equation}
\label{HzFlat}
 \left(\frac{H}{H_0}\right)^2=(1-\alpha)(1+z)^3+\alpha,
\end{equation}
with $\alpha$ being the only free parameter, besides $H_0$, just like on the standard flat 
$\Lambda$CDM model. Note that now $\Omega_{meff}= 1 - \alpha$.

\section{Transition redshift and Supernova bounds}

To begin with, we first observe that by combining Eqs. (\ref{fried}) and (\ref{frw_p}), we have
\begin{equation}
 \frac{\ddot{R}}{R}=-\frac{4\pi G}{3}(\rho_{bar} + \rho_{dm} +3p_c)
\end{equation}

Given that $p_c=-\alpha\rho_{c0}$, one may find:
\begin{equation}
 \frac{\ddot{R}}{R}=-\frac{4\pi G}{3\rho_{c0}}\left[(\Omega_{bar} + \Omega_{dm}- \alpha)(1+z)^3-2\alpha\right]
\end{equation}

When this expression vanishes, one may find the following expression for the transition redshift:
\begin{equation}\label{zt}
 z_t=\left(\frac{2\alpha}{\Omega_m-\alpha}\right)^{1/3}-1.
\end{equation}
Naturally, in order to estimate  the transition redshift it is necessary to constrain the value of $\alpha$ from observations.

Let us now discuss the constraints from distant type Ia SNe data  on the class of CCDM accelerating cosmologies proposed here. 
In what follows we consider both the curved and flat scenarios. In principle, since $H_0$ can be determined from the Hubble Law, the model has only two
independent parameters, namely, $\alpha$ and $\Omega_m$ or, equivalently,  $\Omega_{meff}$ (see Eq. (\ref{Hz}) for $H(z)$). 
However, in the following  analyzes we marginalize over the Hubble parameter. 


The predicted distance modulus for a supernova at redshift $z$, given a set of
parameters $\mathbf{s}$, is
\begin{equation} \label{dm}
\mu_p(z|\mathbf{s}) = m - M = 5\,\mbox{log} d_L + 25,
\end{equation}
where $m$ and $M$ are, respectively, the apparent and  absolute
magnitudes, the complete set of parameters is $\mathbf{s} \equiv
(H_0, \alpha, \Omega_m)$, and $d_L$ stands for the luminosity distance (in
units of megaparsecs),
\begin{equation}
d_L = c(1 + z)\int_{0}^{z} {\frac{dz'}{{H}(z';\mathbf{s})}},
\end{equation}
with $z'$ being a convenient
integration variable, and ${H}(z; \mathbf{s})$ the expression given
by Eq. (\ref{Hz}).

\begin{figure*}
\centerline{\epsfig{file=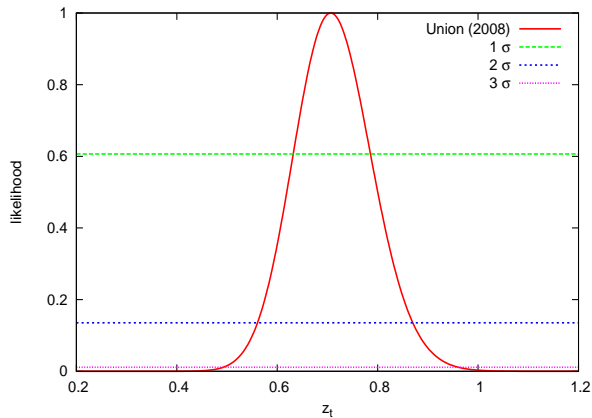,width=80mm,height=58mm}
\hskip
0.1in}\caption{The likelihood of the transition redshift ($z_t$) for the flat CCDM model as given by the Union SNeIa data (2008).The  transition redshift is constrained by  ${z_t=0.706^{+0.080+0.16+0.26}_{-0.074-0.15-0.21}}$.}
\end{figure*}

In order to constrain the free parameters of the model consider now the recent sample (Union 2008),  containing 307 Supernovas as published by  Kowalski and coworkers \cite{Union08}.
The best fit to the set of parameters $\mathbf{s}$ can be estimated by
using a $\chi^{2}$ statistics with
\begin{equation}
\chi^{2} = \sum_{i=1}^{N}{\frac{\left[\mu_p^{i}(z|\mathbf{s}) -
\mu_o^{i}(z)\right]^{2}}{\sigma_i^{2}}},
\end{equation}
where $\mu_p^{i}(z|\mathbf{s})$ is given by Eq. (\ref{dm}),
$\mu_o^{i}(z)$ is the extinction corrected distance modulus for a
given SNe Ia at $z_i$, and $\sigma_i$ is the uncertainty in the
individual distance moduli. In the joint analysis, by marginalizing on the nuisance parameter $h$ ($H_0 = 100h$ km s$^{-1}$ Mpc$^{-1}$) we find $\alpha=0.93^{+0.22+0.35+0.46}_{-0.26-0.44-0.63} $ and $\Omega_m = 1.34^{+0.34+0.54+0.72}_{-0.40-0.68-0.98}$ at
$68.3\%$, $95.4\%$ and $99.7\%$ of confidence level, respectively, with $\chi^{2}_{min}=310.23$ and $\nu=305$ degrees of freedom. The reduced $\chi^{2}_{r}=1.017$ where ($\chi^{2}_{r}=\chi^{2}_{min} /\nu$), thereby showing that the model provides a very good fit to these data and that a closed Universe dominated only by CDM and Baryons is favored by these data.  

In figure 1a we display the space parameter $\Omega_{meff} - \alpha$. Note that the best fit for such a quantity (the effectively measured density parameter) is $\Omega_{meff} = \Omega_m - \alpha = 0.41^{+0.13+0.21+0.29}_{-0.15-0.26-0.37}$.

In the flat case, the only free parameter is $\alpha$, and, as it should be expected, the model also provides a good fit to the SN data. In Figure 1b,  we show the likelihood function for $\alpha$, given by $L\propto e^{-\chi^2/2}$. In this analysis, we find $\alpha=0.713^{+0.027+0.052+0.077}_{-0.028-0.058-0.089}$, for $\chi^2_{min}=311.94$ and $\chi^2_r=1.019$, for 306 degrees of freedom. This is also an extremely good fit thereby showing that we can fit the SNe Ia data with only pressureless matter and creation of CDM particles.

The value of the transition redshift is implicitly  dependent on the curvature parameter. For the curved CCDM model, by inserting the best-fits of $\alpha$ and $\Omega_m$  into $(\ref{zt})$ one obtains the central value $z_t =0.65$ whereas for the flat scenario the transition redshift is a little higher ($z_t = 0.71$) which is in accordance to the fact that we have more matter in the former case. In Figure 2 we display the likelihood for the flat case.

\section{Conclusion}

A new creation cold dark matter (CCDM) cosmology has been proposed. In this late time CDM dominated model, the  vacuum energy density  parameter is  $\Omega_{\Lambda} = 0$, and, therefore, the so-called cosmological constant problem \cite{weinb, decaying} is absent.  The late time acceleration is  powered here by an irreversible creation of CDM particles and the value of  $H_0$ does not need to be small in order to solve the age problem. 

\begin{table}[ht]
 \caption{$\Lambda$CDM vs. CCDM}
\centering
\begin{tabular}{c c}
\hline\hline
$\Lambda$CDM & CCDM \\ [0.5ex]
\hline
$\Omega_\Lambda$ & $\alpha$ \\ 
$\Omega_m$ & $\Omega_{meff}\equiv \Omega_m - \alpha$ \\ 
Vacuum DE & Creation of CDM \\ 
Acceleration ($z_t \approx 0.71$, $k=0$)\, & \,acceleration ($z_t \approx 0.71$, $k=0$) \\ [1ex]
\hline
\end{tabular}
\label{tab1}
\end{table}

It is worth noticing the existence of a dynamic equivalence between CCDM and $\Lambda$CDM cosmologies at the level of the background equations.  Actually, the CCDM scenario can formally be interpreted as a two component fluid mixture: a pressureless matter with density parameter, $\Omega_{eff} = \Omega_m -\alpha$, plus a ``vacuum fluid"  with $\rho_v=-p_v = \alpha \rho_{co}$, where $\rho_{co}$ is the critical density parameter.  A simple qualitative  comparison between both approaches is summarized on Table 1. Note that for nonflat CCDM, there are two dynamic free parameters, namely, $\alpha$ and  $\Omega_m$ (or $\Omega_{meff}$), similarly to nonflat $\Lambda$CDM,  whose dynamic free parameters are  $\Omega_{\Lambda}$ and $\Omega_m$. For the flat CCDM case,  there is just one dynamic free parameter (like in  flat $\Lambda$CDM model), say, $\alpha$.  This formal equivalence explains why CCDM scenarios provide  an excellent fit to the observed dimming of distant type Ia supernovae (see  text and Figs. 1a and 1b). As it appears, one may say that $\Lambda$CDM cosmology is one of the possible effective descriptions of cold dark matter creation scenarios.    

On the other hand, since the creation mechanism adopted here is  classically described as an irreversible process \cite{Prigogine,LCW}, the basic  problem with this new cosmology is related  to  the absence of a consistent approach based in quantum field theory in curved spacetimes.  However,  the  last 30 years thinking about the cosmological constant problem are suggesting that the possible difficulties in searching for a more rigorous quantum approach for matter creation in  the expanding Universe are much smaller than in the so-called $\Lambda$-problem.
Indeed, the basic tools have already been discussed long ago \cite{Parker,BirrellD}, and, as such, the problem now is reduced  to take into account properly the entropy production rate present in the creation mechanism and the associated creation pressure. 

Finally, for those believing that  $\Lambda$CDM contains all the physics that will be needed to confront  the next generation of cosmological tests we call attention for the model proposed here. It is simple like $\Lambda$CDM, has the same dynamics, and, more important, it is based only in cold dark matter whose status is relatively higher than any kind of dark energy. Naturally, new constraints on the relevant parameters ($\alpha$ and $\Omega_m$) from complementary observations need to be investigated in order to see whether the CCDM model proposed here provides a realistic description of the observed Universe.  In principle, additional tests  measuring  the matter power spectrum, the weak gravitational lensing distortion by foreground galaxies and the cluster mass function may decide between $\Lambda$CDM and CCDM cosmologies. New bounds on the CCDM parameters coming from background and perturbed cosmological equations will be discussed in a forthcoming communication.

\begin{acknowledgments}
JASL would like to thank Gary Steigman for helpful discussions and by the warm hospitality during the visit to CCAPP at the Ohio State University.  JFJ is supported by CNPq, FAO is supported by CNPq (Brazilian Research Agencies) and JASL is partially supported by CNPq and FAPESP under grants 304792/2003-9 and 04/13668-0, respectively.
\end{acknowledgments}

\end{document}